\documentclass{aa} 

\usepackage{graphicx}
\usepackage{txfonts}

\newcommand{\speed}[1]{#1 km~s${}^{-1}$}
\newcommand{\accel}[1]{#1 m~s${}^{-2}$}
\newcommand{\nfig}[1]{Figure~\ref{#1}}
\newcommand{\rsun}[1]{${#1}\,R_\odot$}

\begin{document}
\title{White-light QFP Wave Train and the Associated Failed Breakout Eruption}
\author{Yuandeng Shen\inst{1,2,3,5}
 \and
 Surui Yao\inst{1}
 \and
 Zehao Tang\inst{1,3}
 \and
 Xinping Zhou\inst{1,3}
 \and
 Zhining Qu\inst{4}
 \and
 Yadan Duan\inst{1,3}
 \and
 Chengrui Zhou\inst{1,3}
 \and
 Song Tan\inst{1,3}}

\institute{Yunnan Observatories, Chinese Academy of Sciences,  Kunming, 650216, China \email{ydshen@ynao.ac.cn}
\and
State Key Laboratory of Space Weather, Chinese Academy of Sciences, Beijing 100190, China
\and
University of Chinese Academy of Sciences, Beijing 100049, China
\and
Department of Physics and Electronic Engineering, Sichuan Normal University, Chengdu 610101, China
\and
Yunnan Key Laboratory of Solar Physics and Space Science, Kunming, 650216, China}

\date{Received May 2, 2022; accepted July 16, 2022}

 
  \abstract
   {Quasi-periodic fast-propagating (QFP) magnetosonic wave trains are commonly observed in the low corona at extreme ultraviolet wavelength bands. Here, we report the first white-light imaging observation of a QFP wave train propagating outwardly in the outer corona ranging from 2 to \rsun{4}. The wave train was recorded by the Large Angle Spectroscopic Coronagraph on board the {\em Solar and Heliospheric Observatory}, and it was associated with a {\em GOES} M1.5 flare in NOAA active region AR12172 at the southwest limb of the solar disk. Measurements show that the speed and period of the wave train were about \speed{218} and 26 minutes, respectively. The extreme ultraviolet imaging observations taken by the Atmospheric Imaging Assembly on board the {\em Solar Dynamic Observatory} reveals that in the low corona the QFP wave train was associated with the failed eruption of a breakout magnetic system consisting of three low-lying closed loop systems enclosed by a high-lying large-scale one. Data analysis results show that the failed eruption of the breakout magnetic system was mainly because of the magnetic reconnection occurred between the two sided low-lying closed-loop systems. This reconnection enhances the confinement capacity of the magnetic breakout system because the upward-moving reconnected loops continuously feed new magnetic fluxes to the high-lying large-scale loop system. For the generation of the QFP wave train, we propose that it could be excited by the intermittent energy pulses released by the quasi-periodic generation, rapid stretching and expansion of the upward-moving, strongly bent reconnected loops.}


   \maketitle
%

\section{Introduction}
Coronal quasi-periodic fast-propagating (QFP) wave trains observed at extreme ultraviolet wavelengths are composed of multiple coherent and concentric wavefronts, emanating successively close to the epicenter of flares, and propagate at supersonic speeds from several hundred to a few thousand \speed{} \citep[e.g.,][]{2011ApJ...736L..13L, 2012ApJ...753...53S, 2018ApJ...853....1S,2014SoPh..289.3233L, 2018ApJ...868L..33L, 2022SoPh..297...20S}. In addition, some possible signals of QFP wave trains are often recorded in radio observations \citep[e.g.,][]{2018ApJ...861...33K,2013A&A...550A...1K,2018ApJ...855L..29K}. Generally, a QFP wave train is typically associated with an impulsively generated broadband perturbation such as a flare, and its propagation often experiences three distinct phase including periodic phase, quasi-periodic phase and decay phase \citep{1983Natur.305..688R,2021MNRAS.505.3505K}. In addition, a QFP wave train often has a characteristic tadpole or boomerangs signature in the time-dependent wavelet power spectrum \citep[][]{2004MNRAS.349..705N, 2021MNRAS.505.3505K}. Observations showed that QFP wave trains typically first appear at a distance greater than 100 Mm, and they can propagate a large distance up to 200 -- 400 Mm from their origins \citep{2014SoPh..289.3233L, 2022SoPh..297...20S}. In particular, theoretical studies suggested that a QFP wave train will accumulate at the a null point on the path due to the refraction effect around the null point, and the wave energy accumulated around the null point is enough to induce magnetic reconnection \citep[e.g.,][]{2004A&A...420.1129M, 2005A&A...435..313M, 2009A&A...493..227M, 2012A&A...545A...9T, 2018SSRv..214...45M}.  According to the statistical study performed by  \cite{2022SoPh..297...20S}, spatially resolved QFP wave trains can be divided into narrow and broad types. Typically, narrow QFP wave trains propagate along magnetic field lines and with a small intensity amplitude of about 1\%--8\% and an angular width of about 10--80 degrees \citep[e.g.,][]{2013A&A...554A.144Y, 2014A&A...569A..12N, 2017ApJ...851...41Q, 2018ApJ...860...54O, 2019ApJ...871L...2M, 2021ApJ...908L..37M, 2022ApJ...926L..39D}, while broad QFP wave trains propagate across magnetic field lines parallel to solar surface and with a large intensity amplitude of about 10\%--35\% and an large angular width of 90--360 degrees \citep{2012ApJ...753...52L, 2017ApJ...844..149K, 2019ApJ...873...22S, 2021SoPh..296..169Z, 2022ApJ...930L...5Z}. In observations, narrow QFP wave trains mainly appear at the 171 \AA\ (occasionally at 193 and 211 \AA\ channels, see \cite{2013SoPh..288..585S} and \cite{2016AIPC.1720d0010L}) channel of the Atmospheric Imaging Assembly \citep[AIA;][]{2012SoPh..275...17L} on board the {\em Solar Dynamic Observatory} ({\em SDO}), while broad QFP wave trains can be observed in  the AIA's all EUV channels. These differences suggest that narrow and broad QFP wave trains might have different origins and physical properties.

There are two main competing physical mechanisms for the generation of QFP wave trains: the dispersion evolution mechanism and pulsed energy excitation mechanism \citep[see][and references therein]{2022SoPh..297...20S}. The former refer to the scenario that a QFP wave train can be generated through the dispersive evolution of an impulsive broadband perturbation, whose periodicity is determined by the physical properties of the waveguide and the surrounding background \citep[e.g.,][]{1984ApJ...279..857R,2004MNRAS.349..705N,2013A&A...560A..97P,2014A&A...568A..20P}. The latter refer to the mechanism that a QFP wave train is excited by pulsed energy release in the magnetic reconnection, and the periodicity is determined by the wave source \citep[e.g.,][]{2011ApJ...740L..33O,2015ApJ...800..111Y,2016ApJ...823..150T,2021ApJ...911L...8W}. In addition, the leakage of photospheric and chromospheric oscillations into the corona are also proposed as a possible excitation source for QFP wave trains \citep{2003ApJ...599..626B, 2012ApJ...753...53S}. For broad QFP wave trains, previous studies suggested that their generation might tightly associated with nonlinear physical process in magnetic reconnections \citep{2012ApJ...753...52L, 2021SoPh..296..169Z, 2022A&A...659A.164Z}, pulsed energy release caused by unwinding of filament threads \citep{2019ApJ...873...22S}, and the leakage of guided wave trains into the homogeneous quiet-Sun corona \citep{2017ApJ...847L..21P}. To date, it is still an open question about the generation mechanism of QFP wave trains \citep{2022SoPh..297...20S}.

Failed solar eruptions mean that they can not cause interplanetary coronal mass ejections (CMEs) owning to insufficient kinetic energy to overcome the Sun's gravity and downward magnetic forces \citep{2003ApJ...595L.135J, 2011RAA....11..594S}. So far, there are several physical influencing factors have been proposed to explain the failed reason of solar eruptions. For example, the strong overlying magnetic field at low altitude \citep[e.g.,][]{2008ApJ...679L.151L, 2009ApJ...696L..70L, 2018Natur.554..211A}, the small gradient of the overlying magnetic field with respect to the height \citep[e.g.,][]{2006PhRvL..96y5002K, 2015ApJ...814..126Z, 2012ApJ...750...12S},  the insufficient energy released in the low corona through flares or weak kinked magnetic flux ropes \citep{2005ApJ...630L..97T, 2011RAA....11..594S,2018ApJ...858..121L}, as well as the angle between erupting magnetic flux ropes with respect to the overlying magnetic fields \citep[e.g.,][]{2018ApJ...853..105B, 2019ApJ...877L..28Z}. It should be pointed out that for a specific failed solar eruption, it should be determined by the combination of multiple physical factors, rather than a certain one. In previous studies, to the best of our knowledge, QFP wave trains related to failed breakout eruptions have not yet been reported, and the direct white-light imaging of QFP wave trains in the outer corona up to several solar radii is also very scarce. We note that \cite{1997ApJ...491L.111O} reported propagating quasi-periodic variations in the polarized brightness at heliocentric distances from 1.9 to 2.45 solar radii with a period of about 6 minutes and lifetime in the range of 20--50 minutes, and the authors proposed that these signatures are possibly result from density fluctuations caused by compressional, slow magentosonic waves propagating in polar coronal holes.

In this letter, we report the first white-light observations of a QFP wave train on 2014 October 2 recorded by the Large Angle Spectroscopic Coronagraph \citep[LASCO;][]{1995SoPh..162..357B} on board the {\em Solar and Heliospheric Observatory} ({\em SOHO}), which was associated with a failed breakout eruption and an M1.5 {\em GOES} flare in the low corona. In this study, we mainly use the LASCO/C2 and AIA EUV (at 171, 193, and 131 \AA\ channels) images. The LASCO/C2 images the outer corona from 2 to 6 solar radii using the white-light wavelength band, and it takes images with a 12 minute cadence and a pixel size resolution of  {11\arcsec.9}.  The AIA onboard the {\em SDO} takes full-disk images of the corona up to 0.5 solar radii above the solar limb with a 12 second temporal resolution and a pixel size resolution of {0\arcsec.6}. Taking these observations together, we will discuss how the observed broad QFP wave train in the outer corona was related to the failed breakout eruption in the low corona.
\section{Results}
The eruption occurred in active region AR12172 at about 17:00 UT in the low corona, and it was located at the west limb of the solar disk. According to the {\em GOES} soft X-ray 1--8 \AA\ flux, the eruption was accompanied by an M1.5 flare started and peaked at about 17:10 UT and 17:44 UT, respectively. It should be noted that this flare did not cause any CME besides the notable QFP wavefronts. Therefore, this eruption should be a failed solar eruption \citep[e.g.,][]{2003ApJ...595L.135J, 2011RAA....11..594S}. In addition, the M1.5 flare was followed by an M7.3 flare in the nearby active region AR12173 (see \nfig{fig2} (f)). This M7.3 flare caused a partial halo CME in the outer corona, and its start and peak times were about 18:49 UT and 19:01 UT, respectively.

\begin{figure}[b]
\centering
\includegraphics[width=0.46\textwidth]{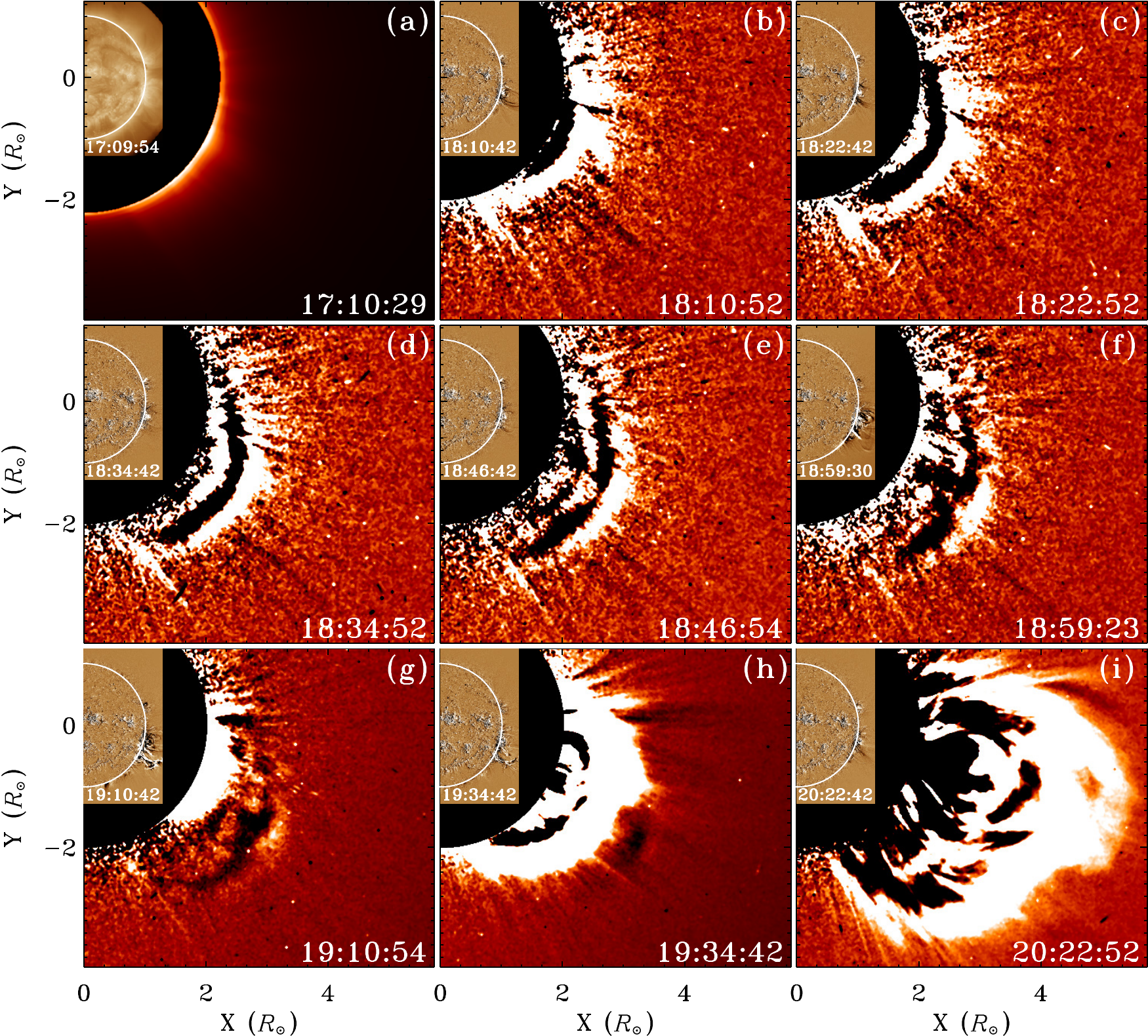}
\caption{Time sequence of composite images made from LASCO/C2 (outer) and AIA 193 \AA\ (inner) observations. Panel (a) is made from direct images at about 18:10 UT before the eruption, while others are made from running-difference images. In each panel, the white circle indicate the disk limb of the Sun, while the black plate represents the inner occulting disk of LASCO/C2. Note that the wave train can be identified in panels (b--f) as coherent white semicircles, while the bright features in panels (g--i) show the CME associated with the following M7.3 flare. }
\label{fig1}
\end{figure}

The pre-eruption corona is shown in \nfig{fig1}(a) with a composite image made from LASCO/C2 and AIA 193 \AA\ direct images at about 17:10 UT, in which coronal loops above the active region can well be identified in the AIA 193 \AA\ image, while the LASCO/C2 image only showed some weak ray-like structures. The QFP wave train is displayed in \nfig{fig1}(b--g); these images show the morphology and evolution of the QFP wave train clearly. It can be seen that the first wavefront appeared at about 18:10 UT in the field-of-view (FOV) of the LASCO/C2 with an angular width of about 90$^\circ$. Then, the second and third wavefronts appeared behind the first one at about 18:22 UT and 18:46 UT, respectively. As shown in \nfig{fig1}(g), due to the appearance of the following CME associated with the M7.3 in the FOV of LASCO/C2, the wave train became too faint to be identified. After that, one can only see the bright CME, but the propagating QFP wave train can not be traced anymore (see \nfig{fig1}(h and i)). Since the distribution of coronal magnetic field in the outer corona is mainly radial, the propagation of the observed QFP wave train should be along the magnetic field. Therefore, according to the definition in \cite{2022SoPh..297...20S}, the present QFP wave train belongs the narrow category.

\begin{figure}
\centering
\includegraphics[width=0.46\textwidth]{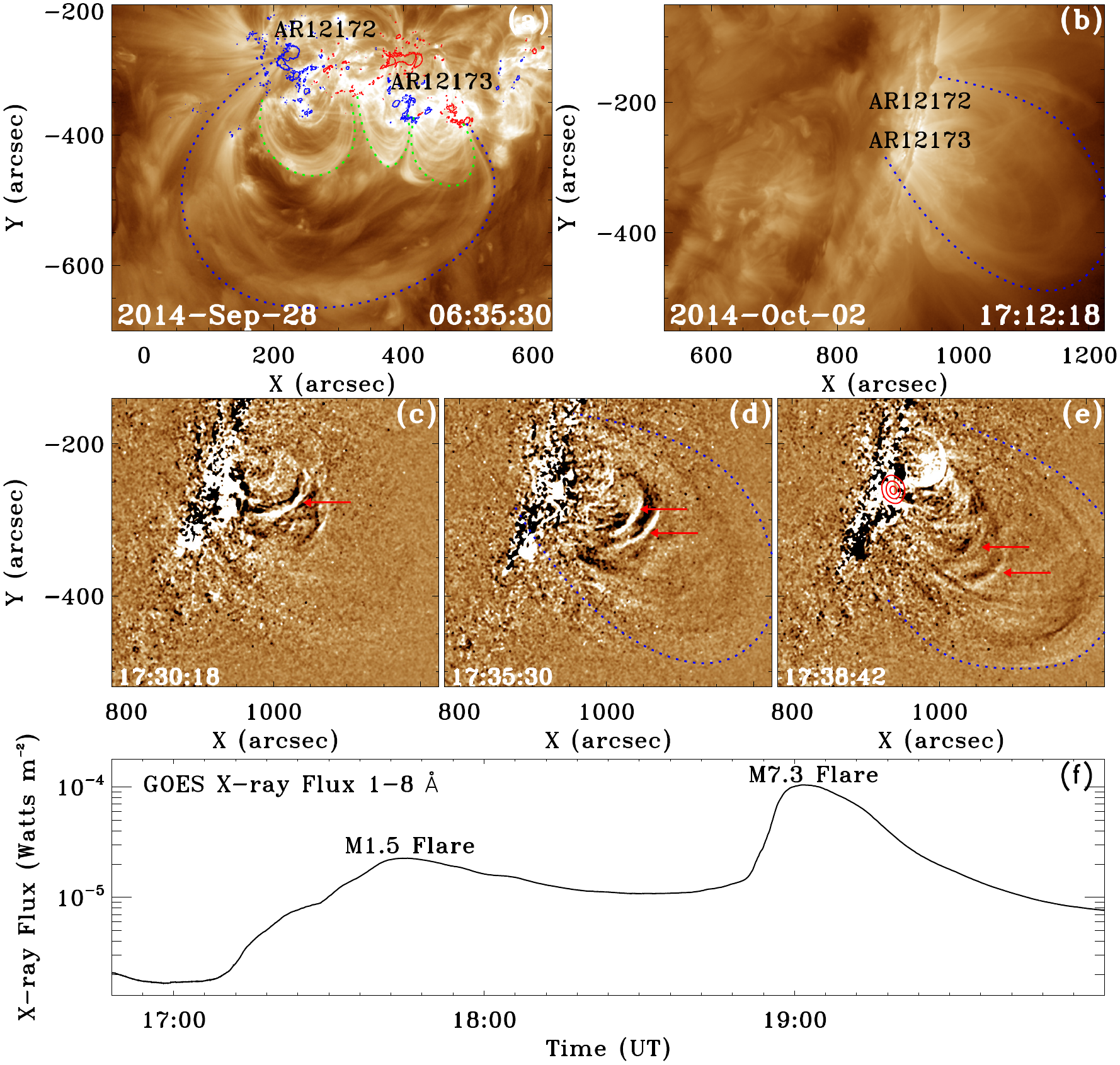}
\caption{Panel (a) is an AIA 193 \AA\ image on 2014 September 28 when AR12172 was on the solar disk, in which the red and blue counters indicate of the  positive and negative magnetic polarities, respectively. Panel (b) is an AIA 193 \AA\ image before the eruption. Panels (c--e) show the eruption process of the event, and the red circles in panel (e) show the contours of the {\em RHESSI} hard X-ray source in the 12--25 Kev energy band. Panel (f) shows the soft X-ray flux record by the {\em GOES} in the energy band of 1 -- 8 \AA, in which the two flares are indicated. }
\label{fig2}
\end{figure}

In the low corona, active regions AR12172 and AR12173 were on the disk limb, and some low-lying arcades are enclosed by a high-lying large loop system which has been outlined with a dotted blue curve in \nfig{fig2} (b). To make clear the magnetic structure of the eruption source region, we checked the active regions on 2014 September 28 when they were close to the disk center (see \nfig{fig2} (a)). It is clear that the eruption source region was a magnetic breakout configuration consisting of three low-lying loop systems (indicated by dotted green curves) enclosed by a large high-lying loop system (indicated by the dotted blue curve). In the framework of magnetic breakout configuration, there exists a null point between the middle low-lying loop and the high-lying large one \citep{1999ApJ...510..485A}, and magnetic reconnection between the two groups of loops often trigger spectacular CMEs in the heliosphere \citep[e.g.,][]{2012ApJ...750...12S,2016ApJ...820L..37C,2013ApJ...764...87L,2021ApJ...911...33C}.

\begin{figure}
\centering
\includegraphics[width=0.46\textwidth]{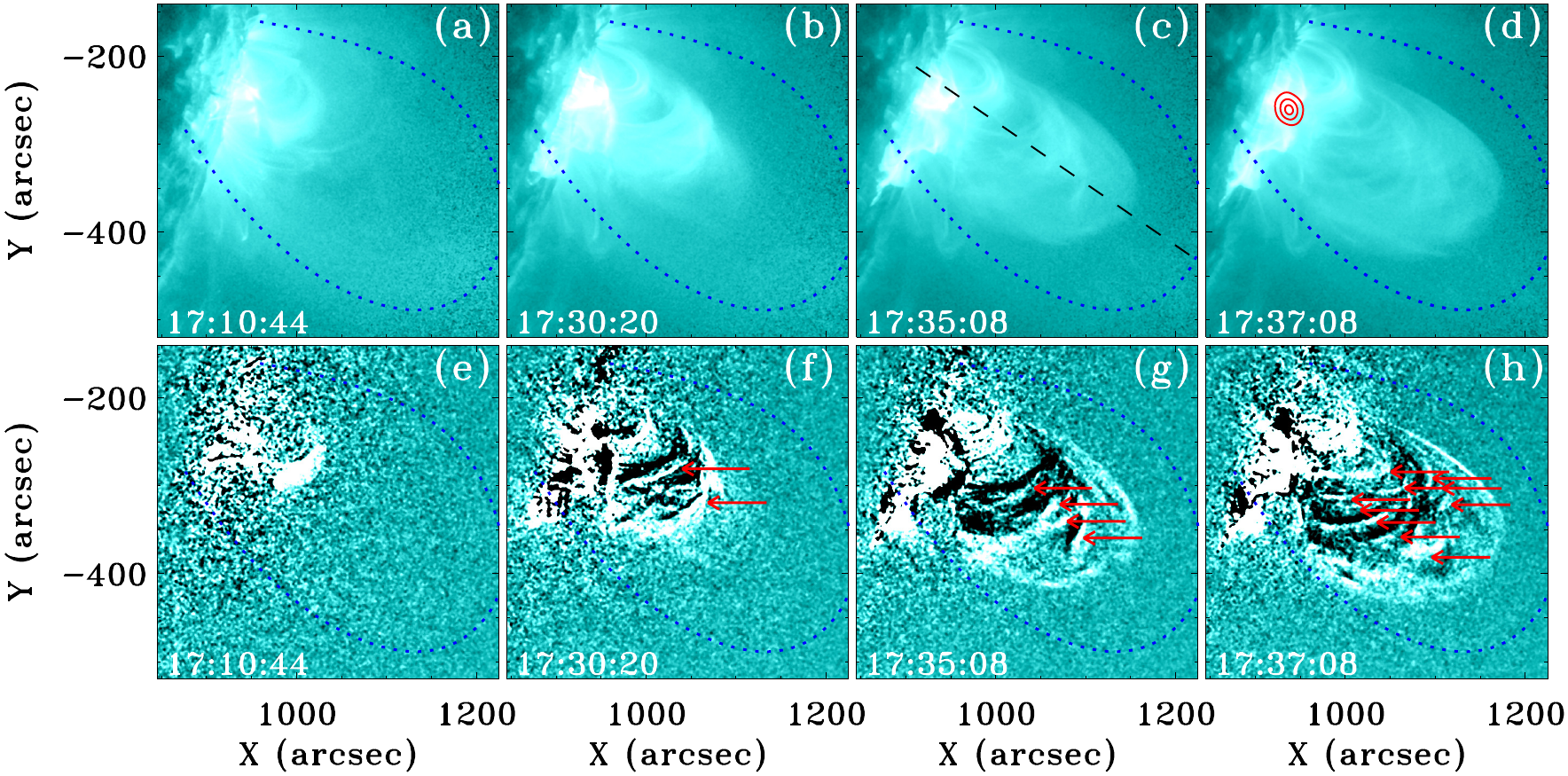}
\caption{AIA 131 \AA\ images show the eruption in the source region. The top row shows the direct time sequence images, while the bottom row shows the corresponding based-difference images. The dashed blue curve in each panel shows the position of the high-lying loop system at 17:12:18 UT (see \nfig{fig2}(b)). The red contours in panel (d) shows the {\em RHESSI} hard X-ray source in the energy band of 12--25 Kev. The red arrows in panels (f)--(h) indicate the outward expanding reconnected loops.}
 \label{fig3}
\end{figure}

The middle row of \nfig{fig2} shows the eruption details of the eruption in the AIA 193 \AA\ running-difference images. As indicated by the red arrows, a chain of bright loops successively generated and expanded outwardly in time, and a hard X-ray source in the 12--25 Kev energy band was detected below these moving loops (see the red contours in \nfig{fig2} (e)). Interestingly, the high-lying large loop system did not erupt but only showed some vertical oscillations during the eruption. These observations suggest that the present event should be a failed one, since we do not find corresponding CME in the outer corona (see the animation available in the online journal). The hotter AIA 131 \AA\ showed the hot eruption core field better than in other channels, but the cooler high-lying large loop system did not appear at this wavelength band (see \nfig{fig3}). The position of the high-lying large loop system determined from the AIA 193 \AA\ at 17:12:18 UT is overlaid as dotted blue curves in each panel in \nfig{fig3}. It can be seen that the hard X-ray source was above the bright post-flare-loop, and hot loops were generated and expanded outward sequentially (see the red contour in \nfig{fig3} (d) and red arrows in the bottom row of \nfig{fig3}). During the eruption period, the expansion of the newly formed hot loops did not exceed the height of the high-lying loop system except for some vertical oscillations (see the animation available in the online journal).

\begin{figure}
\centering
\includegraphics[width=0.46\textwidth]{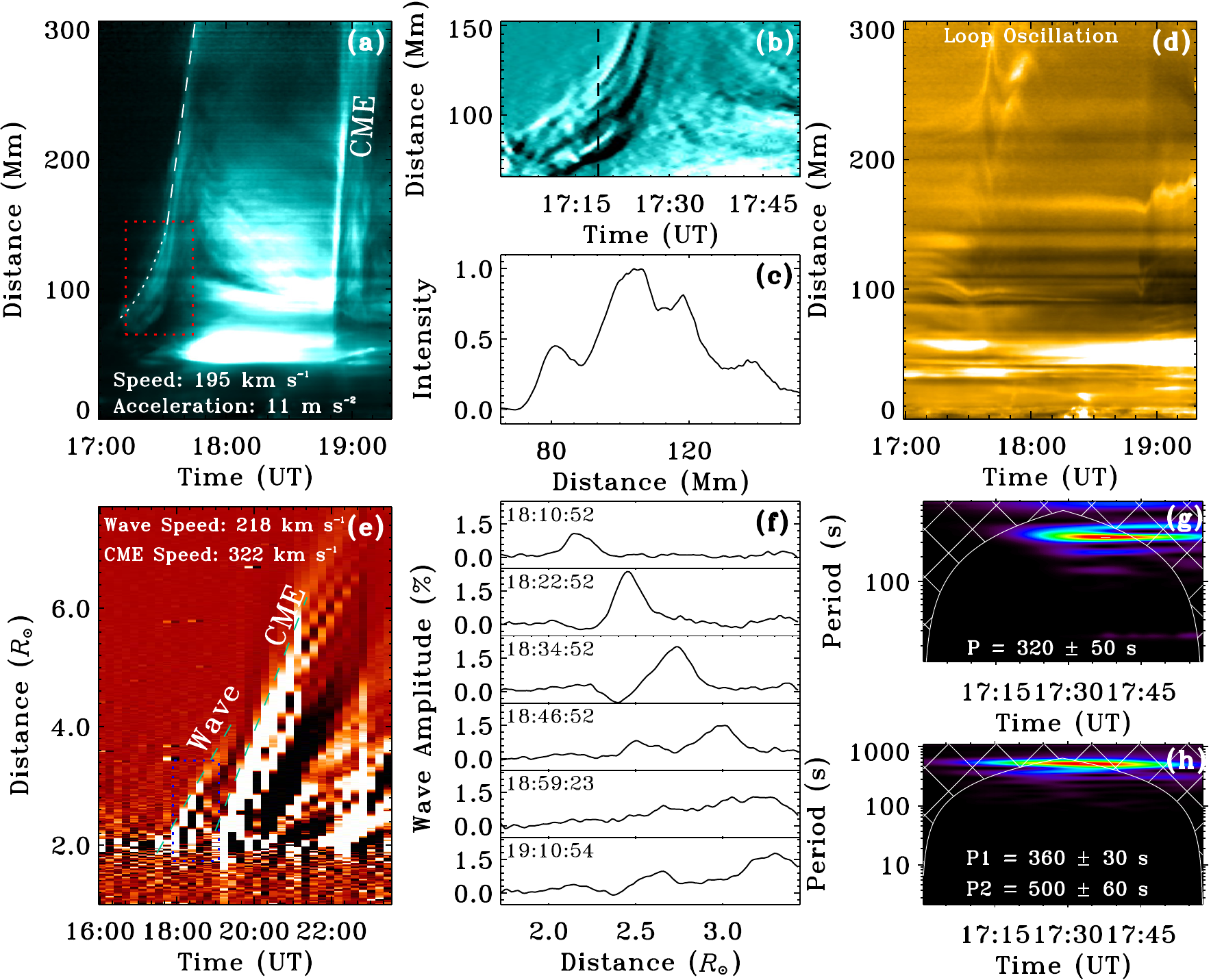}
\caption{Panel (a) is the time-distance plot made from AIA 131 \AA\ images along the black dashed line as shown in \nfig{fig3}(c). Panel (b) shows the time-distance plot within the red box region in panel (a), which is made from the AIA 131 \AA\ running-difference images. The running-difference intensity profile along the dashed black line in panel (b) is plotted in panel (c). Panel (d) is the same with (a), but it is made from AIA 171 \AA\ images. Panel (e) is the time-distance made from LASCO/C2 running-difference images along the same path as panel (a), and panel (f) show the intensity profiles at difference times within the blue dotted box in panel (e). Panel (g) is the wavelet spectrum generated by using the intensity profile as shown in panel (c), while panel (h) is generated by using the time derivative of the {\em GOES} 1--8 \AA\ flux.}
 \label{fig4}
\end{figure}

Time-distance plots are made along the black dashed line as shown in \nfig{fig3} (c) to investigate the kinematics of the expanding loops (\nfig{fig4} (a)), the high-lying large loop (\nfig{fig4} (d)), and the wavefronts (\nfig{fig4} (e)). The newly formed expanding loops experienced an acceleration phase of about 20 minutes from 17:10 UT to 17:30 UT, then they started a linear expansion phase. It is measured that the acceleration and the expanding speed of the expanding loops were about \accel{11} and \speed{195}, respectively. The details of the acceleration phase is displayed in \nfig{fig4}(b) using the time-distance plot made from running-difference images, in which one can see several strips that represent the expanding loops. The corresponding intensity profile along the black line in \nfig{fig4}(b) is plotted in \nfig{fig4}(c), which shows the loops more clearly (each peak represents a loop). \nfig{fig4}(d) shows the time-distance plot made from AIA 171 \AA\ images, which shows the oscillation of the high-lying large loop system, but the hot expanding loops can not be identified. \nfig{fig4}(e) is the time-distance plot made from LASCO/C2 running-difference images, from which it is measured that the speeds of the wave train and the following CME were about \speed{218 and 322}, respectively. The corresponding intensity profiles within the dashed blue box in \nfig{fig4}(e) at different times are plotted in \nfig{fig4}(f). One can see that the amplitude of the wavefronts showed a trend of first increasing and then decreasing. Such a character might suggest the true wave nature of the observed wave train, resembling the evolution pattern of large-scale EUV waves \citep[e.g.,][]{2012ApJ...752L..23S,2015LRSP...12....3W}. Based on the intensity profile at 18:46:52 UT, the wavelength of the wave train can be estimated to be about 345 Mm. By dividing the wavelength by the speed, we can estimate the period of the observed wave train was about 26 minutes. It should be pointed out here that the measured period of the wave train might be very inaccurate due to the significantly low spatiotemporal resolution of the LASCO/C2 observations (12 minutes). With the aid of the wavelet software \citep{1998BAMS...79...61T}, the wavelet spectrums of the expanding loop and the flare pulsations are respectively generated by using the percentage intensity profile as shown in \nfig{fig4} (c) and the time derivative of the {\em GOES} 1--8 \AA\ flux, and they reveal that the period of the expanding loops was about $320 \pm 50$ seconds (see \nfig{fig4} (g)), while that of the accompanying flare were about $360 \pm 30$ and $500 \pm 60$ seconds (see \nfig{fig4} (h)). It is clear that the periods of the flare and the expanding loops in the low corona are significantly shorter than that of the wave train derived from the low resolution LASCO/C2 observations.

\section{Interpretation and Discussions}
Our observational results suggest that the eruption of the source region should be a failed breakout eruption, in which the magnetic reconnection took place between the two low-lying sided loop systems. In this case, the reconnection around the null point results in a negative feedback to the instability of the  breakout magnetic system. Although the failed eruption did not cause a CME in the outer corona,  it excited a QFP wave train due to the rapid stretching and expansion of the newly formed large-scale high-lying loops. 

\begin{figure}
\centering
\includegraphics[width=0.46\textwidth]{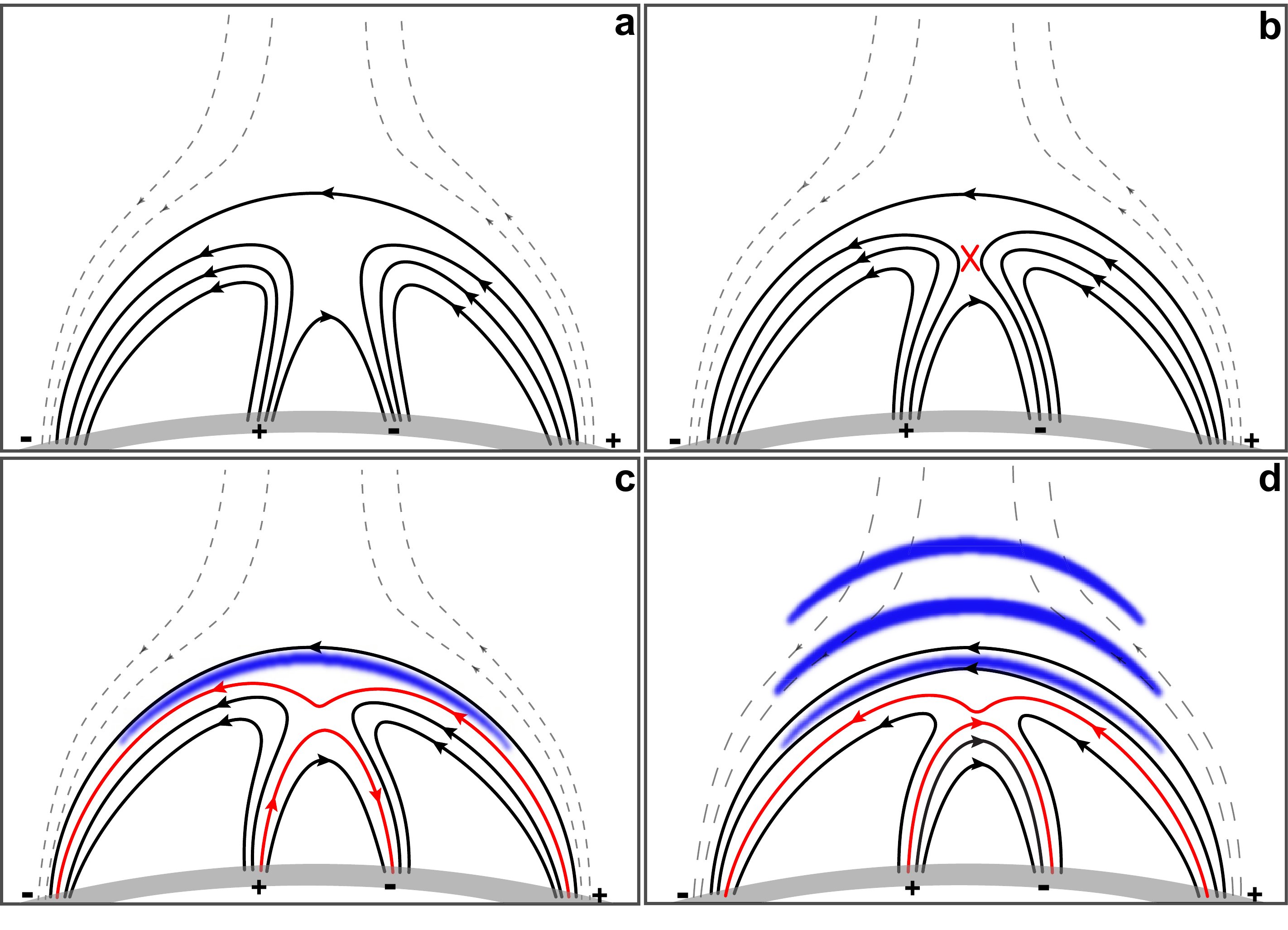}
\caption{A cartoon model demonstrates the failed eruption of the magnetic breakout system and the generation of the QFP wave train. In this figure, only some representative field lines are plotted, and the dashed lines represent the background coronal field. The arrow marked on each line indicates the direction of the magnetic field, the gray thick curve indicates the solar surface, minus and plus signs indicate the negative and positive polarities, respectively. Panel (a) shows the initial magnetic configuration. The red cross in panel (b) indicates the reconnection site, while the red curves in panels (c) and (d) represent the newly formed reconnected field lines. The blue thick curves in panel (c) and (d) show the wavefronts generated by the stretching and expansion of the high-lying reconnected coronal loops.}
 \label{fig5}
\end{figure}

We draw a cartoon in \nfig{fig5} to illustrate the proposed physical picture. In the cartoon, only some representative magnetic field lines are plotted, the minus and plus signs indicate respectively the negative and positive polarities, and the dashed curves represent the background open field lines. \nfig{fig5}(a) shows the initial magnetic configuration, which is a typical magnetic breakout topology consisting of three low-lying closed loop systems enclosed by a large high-lying one. Naturally, a magnetic null point forms between the middle and the high-lying loops (or between the two sided low-lying loops). Due to some unknown reasons, the two sided low-lying loops move close to each other and therefore trigger magnetic reconnection between them (see the red cross symbol in \nfig{fig5}(b)). The consequence of the reconnection will cause the formation of two new loop systems as shown by the red curves in \nfig{fig5}(c). The hard X-ray source should be located around the reconnection site in comparison with the observation. While the downward moving hot reconnected loops form the observed post-flare-loop, the upward moving reconnected loops correspond to the observed expanding loops in the hot AIA 131 \AA\ images. Since the middle section of the upward moving reconnected loop is curved upwards, upward magnetic tension force will stretch it and lead to the outward acceleration and expansion. When the loop becomes curved downward, the magnetic tension force also becomes downward like the initial high-lying large loop. As a result, the loop will decelerate and move downward to reach a stable state, during which the loop might undergo an oscillation phase. We propose that each upward moving reconnected field line can excite a wavefront during its fast acceleration phase. Due to the sequential generation of upward moving reconnected magnetic field lines in the reconnection, it is reasonable to expect the generation of an outward propagating QFP wave train as shown in \nfig{fig5}(d). Our observation not only revealed the reconnection around the null point, but also the acceleration, expansion, and contraction of the newly formed high-lying loops.

The magnetic breakout mode was initially developed to interpret the generation of large-scale CMEs, in which the magnetic reconnection occurs between the middle low-lying loop and the high-lying one \citep{1999ApJ...510..485A}. In this case, the reconnection rapidly removes the high-lying restraining field and therefore let the low-lying core field to escape. To date, this model has been confirmed by many multi-wavelength observations, and it can well explain multi-ribbon flares in tripolar and quadrupolar magnetic regions  \citep[e.g.,][]{2001ApJ...560.1045S, 2004ApJ...617..589L, 2012ApJ...750...12S, 2016ApJ...820L..37C,2009ApJ...693.1178V,2019ApJ...885L..11S,2021ApJ...911...33C,2021ApJ...923...45Z} and small-scale coronal jets \citep{2017Natur.544..452W, 2018ApJ...852...98W, 2018ApJ...854..155K, 2019ApJ...885L..15K, 2019ApJ...874..146H, 2019ApJ...885L..11S, 2011ApJ...735L..43S, 2021RSPSA.47700217S}. The magnetic breakout mode can also produce failed solar eruptions without association to CMEs, if the reconnection occurs between the two sided low-lying loop systems as shown in our cartoon (see \nfig{fig5}). In this case, the upper reconnected loops spring upward and continuously feed magnetic flux to the previous existing high-lying loop system during the reconnection. Therefore, the confinement ability of the high-lying loop system will be enhanced significantly, and such a negative feedback to the instability of the magnetic breakout system can naturally produce failed solar eruptions. In previous studies, hard X-ray source above failed filament eruptions \citep[e.g.,][]{2003ApJ...595L.135J, 2006ApJ...653..719A} and the presence of four weak flare ribbons preceded a strong two ribbon flare \cite{2005ApJ...628L.163W} might attribute to the magnetic reconnection between the two sided low-lying loop systems. In addition, numerical simulation results also indicate that such a reconnection do cause failed eruptions \citep{2008ApJ...680..740D}. 

Basically, we propose that the wave train was possibly excited by the rapid expansion of the newly formed high-lying loops as illustrated by the cartoon, although the period of the wave train (26 minutes) showed a large difference to those of the expanding loops ($320 \pm 50$ seconds), and the wave train was not detected in the low corona. Here, we propose two possible reasons to reconcile such a discrepancy. Firstly, some relatively weak wavefronts in the QFP wave train can not be detected by the LASCO/C2 due to the low spatiotemporal resolution, or some of the initially generated wavefronts had been dissipated when they reached up to the FOV of the LASCO/C2. Secondly, the LASCO/C2 can not detect periods less than 24 minutes because of the limitation of the Nyquist frequency of 0.7 mHz given by its 12 minutes cadence. For the fact that the QFP wave train did not appear in the FOV of the AIA, we think that it might be due to the small FOV of the AIA. Previous observations indicate that a QFP wave train often firstly appears at a distance greater than 100 Mm from the associate flare epicenter \citep{2022SoPh..297...20S}. However, for our case, such a distance has out of the AIA's FOV. In a numerical simulation, \cite{2017ApJ...847L..21P} found that the geometrical dispersion associated with the waveguide suppresses the nonlinear steepening for trapped compressive perturbations. This might be the reason that QFP wave trains always appear at some certain distance far from the generation source region.

It is noted that one period of the flare ($360 \pm 30$) was similar to that of the expanding loops ($320 \pm 50$ seconds), which might suggest that the intermittent energy release caused by stepwise reconnection between different loops can directly modulate the flare pulsation. The present observation of the the sequential generation of expanding loops provides clear evidence to account the existence of some common periods in QFP wave trains and the accompanying flares. In addition, the generation mechanism of quasi-periodic pulsations in flares is still under a hot debate, although various candidate models have been documented in literature \citep[e.g.,][]{2009SSRv..149..119N,2016SoPh..291.3143V,2020STP.....6a...3K,2021SSRv..217...66Z,2020ApJ...888...53L, 2020ApJ...893....7L}. Our observation also provides an explanation to the generation of quasi-periodic pulsation in flares.

The energy flux carried by the QFP wave train can be estimated from the kinetic energy of the perturbed plasma that propagates at the group speed, i.e., $E=(\frac{1}{2}\rho v_{\rm 1}^2)v_{\rm gr}$, where $\rho$ is the plasma density, $v_{\rm 1}$ is the disturbance amplitude of the locally perturbed plasma, and $v_{\rm gr}$ is the group speed of the wave train. For a rough estimation, one can assume that the group speed is equal to the value of the measured phase speed. In addition, in the optically thin corona, the emission intensity $I$ is directly proportional to the square of the plasma density, i.e., $I \propto \rho ^{2}$. Here, it should be pointed out that the emission intensity is also proportional to the column-depth perturbations, and it is especially pronounced if the spatial resolution is limited \citep{2003A&A...397..765C,2012A&A...543A..12G}. For a rough estimation, the density modulation of the background density $\frac{d\rho}{\rho}$ can be written as $\frac{dI}{2I}$, and the energy flux of the perturbed plasma can be rewritten as $E \geq \frac{1}{8} \rho v_{\rm ph}^{3} (\frac{\rm dI}{I})^{2}$ if one assumes that $\frac{v_{\rm 1}}{v_{\rm ph}}$ is equal or greater than $\frac{\rm d \rho}{\rho}$ \citep[see,][and references therein]{2022SoPh..297...20S}. For the present case, the phase speed and the relative amplitude of the wave train are about \speed{218} and 50\%, respectively. In addition, based on the coronal plasma density model presented in \cite{1999ApJ...523..812S}, the electron number densities at the heights of 2 -- \rsun{4} are 0.5 -- $0.1\times10^6 { \rm~cm}^{-3}$. Therefore, we derived that the energy flux of the wave train at the height of 2 -- \rsun{4} is in the range of 324 -- 65 $\rm erg ~ cm^{-2} {\rm ~s}^{-1}$.

For a fast-mode wave guided by a field-aligned plasma non-uniformity, its speed has a value somewhere between the Alfv\'en speeds inside and outside the non-uniformity. Therefore, assuming the observed wave train was a linear fast-mode magnetosonic wave along coronal magnetic field lines, we can estimate the magnetic field strength in the height range of 2 -- \rsun{4} by using the measured wave speed, i. e., $v_{\rm ph}=v_{\rm A}=\frac{B}{\sqrt{4\pi\rho}}=\frac{B}{\sqrt{4\pi\mu m_{\rm p}1.92n_{\rm e}}}$, where $\mu=1.27$ is the mean molecular weight in the corona, $\rm{m_p} = 1.67\times 10^{-24}$\,g is the proton mass. The derived results show that the magnetic field strength in the height range of 2 -- 4R$_\odot$ was in the range of 0.11 -- 0.05 Gausses.

\section{Conclusions} \label{conclusion}
we report the first white-light imaging observation of a QFP wave train recorded by LASCO/C2 high above the limb (at heliocentric distances from 2 to 4 solar radii), whose speed, period and wavelength were about \speed{218}, 26 minutes and 345 Mm, respectively. We propose that the QFP wave train might be excited by the rapid outward expansion of low coronal loops generated by the magnetic reconnection between the two sided low-lying loop systems of the magnetic breakout configuration in active region AR12172. In addition to the generation of the QFP wave train, the reconnection also led to the enhancing of the high-lying confining magnetic field strength of the system, which can be regarded as a main physical factor to result in a failed solar eruption. We found that the intermittent generation of the outward moving reconnected loops is intimately associated with the stepwise nonlinear reconnection process, which can well account for the existence of common periods in flares and QFP wave trains. Using the wave properties of the wave train and the coronal plasma density model, it is estimated that the energy flux carried by the wave train was in the range of 324 -- 65 $\rm erg ~ cm^{-2}{\rm ~s}^{-1}$, while the magnetic field strength at the height of 2 -- 4R$_\odot$ was in the range of 0.11\,--\,0.05 Gausses.

\begin{acknowledgements}
The authors thank the excellent data provided by {\em SDO} and {\em SOHO} teams, and the anonymous referee for his/her valuable comments to improve the quality of this paper. This work was supported by the Natural Science Foundation of China (12173083, 11922307, 11773068), the Yunnan Science Foundation for Distinguished Young Scholars (202101AV070004), the National Key R\&D Program of China (2019YFA0405000), the Specialized Research Fund for State Key Laboratories, and the Yunnan Key Laboratory of Solar Physics and Space Science (202205AG070009).
\end{acknowledgements}


\end{document}